\def\year{2020}\relax
\documentclass[letterpaper]{article} 
\usepackage{aaai20}  

\usepackage{times}  
\usepackage{helvet}  
\usepackage{courier}  
\usepackage[hyphens]{url}  
\usepackage{graphicx} 
\usepackage{natbib}  
\usepackage{caption} 
\usepackage{algorithm}
\usepackage{algorithmic}
\usepackage{adjustbox} 
\usepackage{booktabs}  
\usepackage{amsmath}   
\usepackage{newfloat}
\usepackage{listings}
\DeclareCaptionStyle{ruled}{labelfont=normalfont,labelsep=colon,strut=off} 
\floatstyle{ruled}
\newfloat{listing}{tb}{lst}{}
\floatname{listing}{Listing}
\usepackage{hyperref}
\hypersetup{
	colorlinks=true,
	urlcolor=magenta,
	citecolor={black}
}

\title{ Efficient Climate Simulation via Machine Learning Method}
\author{Xin Wang,\textsuperscript{\rm 1}
Wei Xue,\textsuperscript{\rm 2}
Yilun Han,\textsuperscript{\rm 1} Guangwen Yang \textsuperscript{\rm 1}\thanks{Corresponding author} \\
\textsuperscript{\rm 1}Department of Computer Science and Technology, Tsinghua University,\\
\textsuperscript{\rm 2}Department of Earth System Science, Tsinghua University\\
x-w19@mails.tsinghua.edu.cn,\\
ygw@tsinghua.edu.cn}

\usepackage{bibentry}

\begin{document}

\maketitle

\begin{abstract}
Hybrid modeling combining data-driven techniques and numerical methods is an emerging and promising research direction for efficient climate simulation. However, previous works lack practical platforms, making developing hybrid modeling a challenging programming problem. Furthermore, the lack of standard data sets and evaluation metrics may hamper researchers from comprehensively comparing various algorithms under a uniform condition. To address these problems, we propose a framework called NeuroClim for hybrid modeling under the real-world scenario, a basic setting to simulate the real climate that we live in. NeuroClim consists of three parts: (1) \textbf{Platform}. We develop a user-friendly platform NeuroGCM for efficiently developing hybrid modeling in climate simulation. (2) \textbf{Dataset}. We provide an open-source dataset for data-driven methods in hybrid modeling. We investigate the characteristics of the data, i.e., heterogeneity and stiffness, which reveals the difficulty of regressing climate simulation data; (3) \textbf{Metrics}. We propose a methodology for quantitatively evaluating hybrid modeling, including the approximation ability of machine learning models and the stability during simulation. We believe that NeuroClim allows researchers to work without high level of climate-related expertise and focus only on machine learning algorithm design, which will accelerate hybrid modeling research in the AI-Climate intersection. The codes and data are released at \url{https://github.com/x-w19/NeuroClim}.
\end{abstract}

\section{Introduction}
\label{sec:introduction}

The anthropogenic climate change and the resulting frequent and intense extreme events have caused widespread impacts on natural and human systems and sometimes led to severe damage and loss~\cite{zhongming2021ar6}. Prediction with climate models provides a critical reference for governments to address climate change~\cite{allan2021ipcc}. 

General circulation models (GCMs) have long been used as climate models for over a half-century and have been applied in climate simulations and projections~\cite{stevens2013climate}. The GCM is a numerical method to simulate large-scale motions in the atmosphere. It uses a grid of 100 $\sim$ 200 kilometers horizontally and dozens of layers vertically. However, the GCM cannot simulate atmospheric fluid processes at smaller scales (i.e., clouds and convection), so it uses a subgrid parameterization based on a limited number of observations and empirical models to represent no atmospheric fluid processes on the grid, which also introduces uncertainties to the GCM~\cite{bony2015clouds,liang2022stiffnessaware,randall2003breaking,emanuel1994large}.

\begin{figure*}[t]
    \centering
    \includegraphics[width=0.7\linewidth]{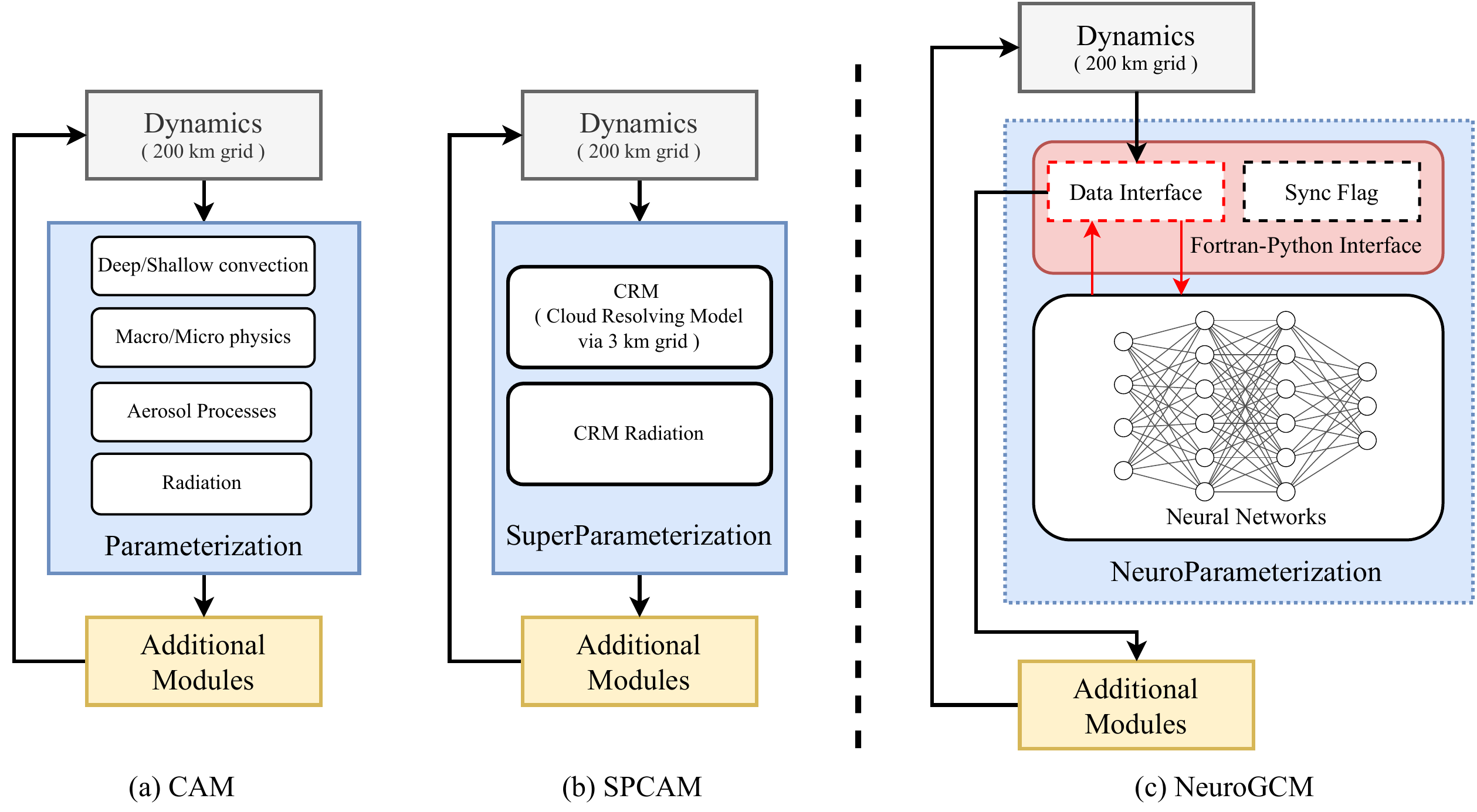}
    \caption{The structures of GCMs. (a) is Community Atmosphere Model (CAM); (b) is super-parameterized CAM (SPCAM); (c) is NeuroGCM (i.e., Hybrid modeling in this work).}
    \label{fig:architecture}
\end{figure*}

The uncertainty of subgrid parameterization makes GCM unable to simulate and capture real climate change, making its climate prediction flawed. The researchers try to use a high-resolution numerical approach instead of the typical subgrid parameterization of the GCM. Specifically, 4km-grid cloud-resolving models (CRMs) are embedded in each 200km grid of the GCM. This approach is also known as Superparameterization~\cite{randall2003breaking,arakawa2004cumulus, khairoutdinov2005simulations}. Superparameterized Community Atmosphere Model (SPCAM) is a commonly used climate model with superparameterization developed by National Center for Atmospheric Research (NCAR). It is based on Community Atmosphere Model (CAM), a widely used climate model. SPCAM uses direct numerical methods to simulate clouds and convection to improve the accuracy of climate predictions. Fig~\ref{fig:architecture}(a) and (b) show the similarities and differences between CAM and SPCAM. However, this introduces additional computational overhead. SPCAM requires $\mathcal{O}(10^{2})$ times more than the typical GCM~\cite{randall2003breaking}. Such high computational demand is problematic for long-term climate simulations.

One possible way to reduce the computational cost is to substitute the tedious cloud-resolving scale numerical simulation with an efficient yet accurate surrogate model. With the recent development of machine learning (ML) technique \cite{liang2020instance,huang2020dianet}, from single-step predictions to simulations on ideal models and even under the real-world setting, important advances (\citet{krasnopolsky2013using,rasp2018deep,wang2021stable} and more summarized in Sec.~\ref{sec:related}) have been made in learning the cloud and convective parameterization schemes for GCMs from data. The basic idea is to combine the numerical process of GCM with machine learning models to design hybrid modeling that fuses the accuracy of high-resolution models (e.g., GCRM or SPCAM) with the efficiency of GCM. First, these methods use simulation results of high-resolution models (e.g., GCRM or SPCAM) as training data. Then the trained model is applied to couple with the coarse-resolution GCMs. Based on the summary in Sec.~\ref{sec:related}, the initial studies show potential in obtaining accurate emulations with an efficient machine learning model. However, the progress in hybrid modeling is still exploratory. The majority of studies~\cite{rasp2018deep,yuval2020stable} based on toy models can examine some idealized climate phenomena but never the realistic, complex climate systems. A few studies~\cite{han2020moist,mooers2021assessing} tried to achieve hybrid modeling in the real-world setting\footnote{\textbf{The real-world setting} is a present-day climate model setting, coupled to a land surface model Community Land Model version 4.0~\cite{oleson2010technical} and forced under prescribed sea surface temperatures and sea ice concentrations~\cite{hurrell2008new}.} but faced difficulty performing actual long-term simulations. These attempts are based on different environments and data and are difficult to compare with each other, as reflected in the following three primary factors. (1) \textbf{Platform.} Hybrid modeling lacks a unified and researcher-friendly experimental platform. (2) \textbf{Dataset.} The hybrid models lack accessible benchmark datasets for a fair comparison. (3) \textbf{Metric.} There lacks a standard evaluation for a fair comparison and systematic research on the stability of prognostic simulation. 

The contribution of NeuroClim is summarized as follows. 

\textbf{Platform.} We develop NeuroGCM, a hybrid modeling platform with data-driven and numerical methods based on the real-world setting. Researchers can design data-driven methods (i.e., machine learning algorithms) and easily deploy them to implement hybrid modeling. The platform is released to the public and readily to use.
    
\textbf{Dataset.} We obtained cloud and radiation data from SPCAM, a $km$-scale resolution enhanced climate model\footnote{SPCAM uses \textbf{3} $km$-scale CRMs to enhance its climate predicting skill and requires $\mathcal{O}(10^{2})$ times more than the \textbf{200} $km$-scale typical GCM~\cite{randall2003breaking}.}, to train machine learning algorithms in hybrid modeling. It is the first open-source hybrid modeling dataset based on the NumPy interface that can serve as a challenging benchmark for comparing various models. We also point out the heterogeneity and stiffness characteristics of this dataset.
    
\textbf{Metric.} We introduce two metrics to verify the performance of the algorithm, including RMSE to evaluate the approximation ability and pseudo-Radiative forcing $\kappa$ to evaluate the simulation stability.

We also provide a reference-level data-driven parameterization scheme that can support long-term stable simulation of NeuroGCM. 
We believe that this work will stimulate both the machine learning community and the climate modeling community to address impactful challenges, such as the interpretability and stability of hybrid modeling and the evaluation of climate prediction techniques for different regions. 

\section{Platform}
\label{sec:platform}

This section introduces NeuroGCM, a hybrid modeling platform in NeuroClim. In NeuroGCM, we creatively use a data-driven parameterization (called NeuroParameterization) instead of the cumbersome SuperParameterization in SPCAM.

\subsection{Overview}
\label{sec:workflow}

Fig.~\ref{fig:architecture}(c) displays the structure of NeuroGCM, which mainly includes dynamics, NeuroParameterization and Python-Fortran Interface. Among them, dynamics is the primary numerical method of GCMs, inherited by NeuroGCM. Dynamics numerically solves the large-scale motions of the atmosphere using grids with a scale of 100-200 km. Specifically, large-scale motions of the atmosphere are governed by hydrodynamic equations in coarse-grid, characterized by slow vertical motion and zero vertical acceleration (more details in Appendix A). Then comes NeuroParameterization, which uses a data-driven approach to predict cloud and convective processes at 4km-scale and obtain the necessary variables at each step in long-term climate simulations.

Data-driven NeuroParameterization and dynamics based on numerical methods are used to solve for atmospheric fluid processes at the corresponding scales, respectively, to complement each other for the entire climate simulation. The PythonFortran Interface supports the data interaction between NeuroParameterization and dynamics and is described in subsection~\ref{sec:FPI}.


\subsection{NeuroParameterization}

In the atmosphere, fluid processes consist of both largescale motions and cloud and convective (i.e., subgrid-scale) processes. The scales of cloud and convective processes range from a few hundred meters to several kilometers. Therefore, these cloud and convective processes are inside the coarse-grid (grid size of 200km) and cannot be resolved by dynamics.

\textbf{Parameterization.} Typical GCMs use subgrid parameterization, relying on limited observations and empirical modeling to represent cloud and convective processes. As a compromise method, it can not accurately resolve cloud and convective processes, which introduces uncertainty. 

\textbf{SuperParameterization.} To improve the subgrid parameterization, researchers propose directly using a 4km ultra-high-resolution cloud-resolving model to simulate cloud and convection processes and embed them in the subgrid of Dynamics. This direct numerical method-based subgrid parameterization, called SuperParameterization, improves accuracy but consumes much computing resources.

\textbf{NeuroParameterization.} In this study, we used a data-driven approach instead of SuperParameterization and named as NeuroParameterization. NeuroParameterization accepts data from each grid of Dynamics. Considering all the grids of each simulation step, the shape of input/output variables will be $144 \times 96 \times 30$, where $144 \times 96$ is the latitude and longitude resolution, and 30 is the vertical resolution. These variables form the input and output of the data-driven approach, detailed in Sec.~\ref{sec:dataset}.

\begin{figure}[t]
    \centering
    \includegraphics[width=1\linewidth]{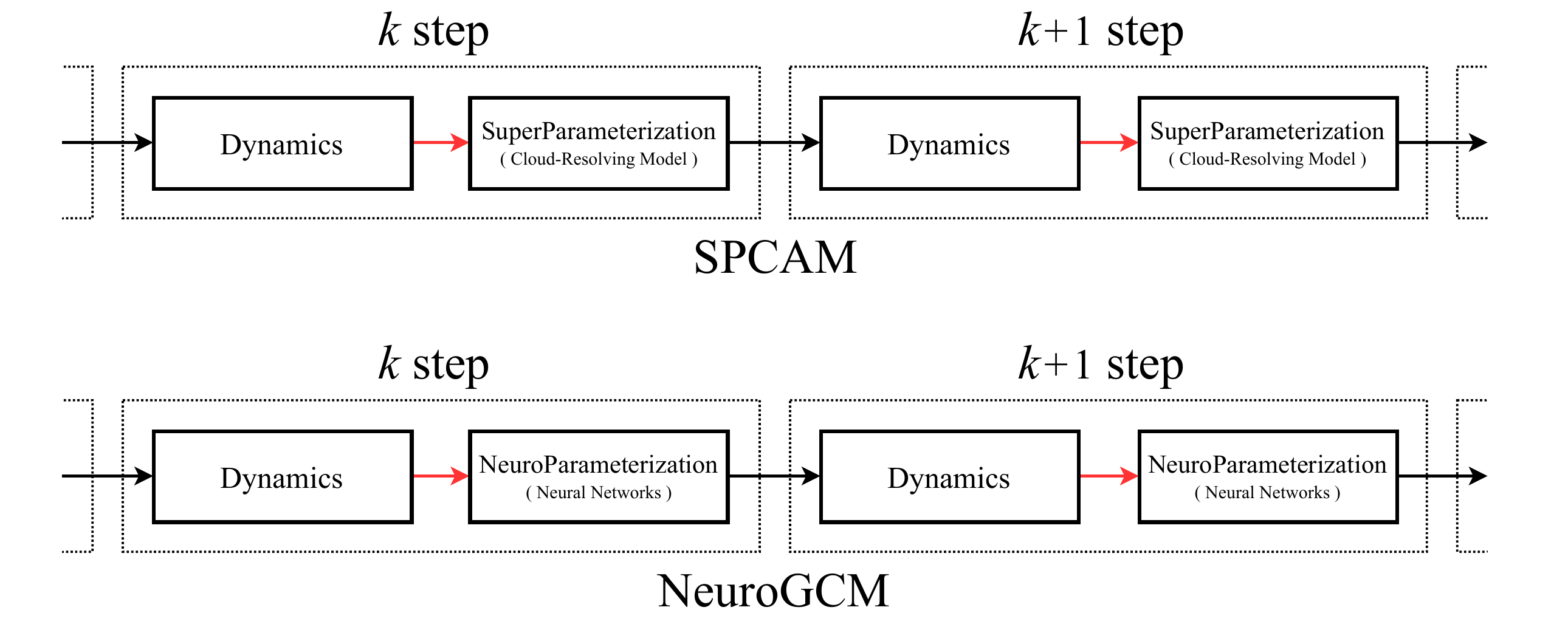}
    \caption{The integration processes of SPCAM and NeuroGCM during simulation.}
    \label{fig:integration}
\end{figure}

\begin{figure*}[t]
    \centering
    \includegraphics[width=0.9\linewidth]{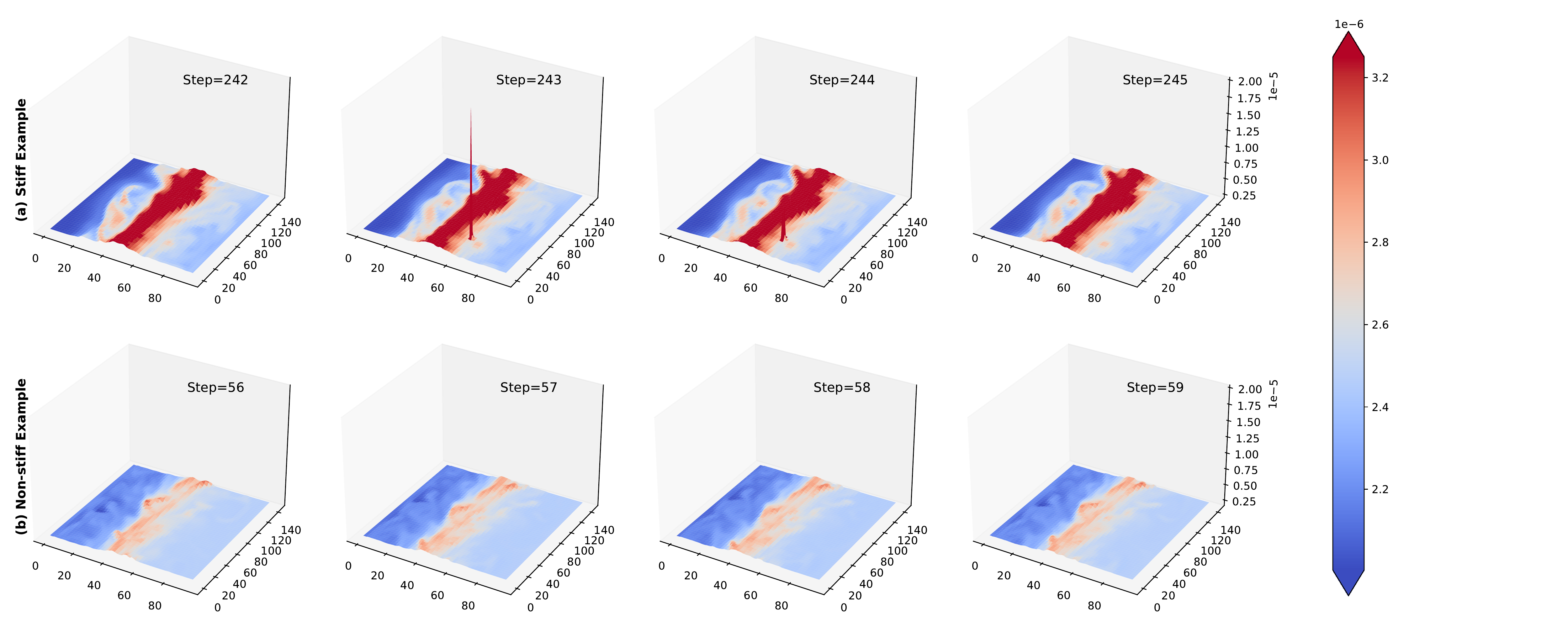}
    \caption{The stiff and non-stiff example. Taking the region of longitude (0$^{\circ} E\sim$ 140$^{\circ} E$) and latitude (0$^{\circ} N\sim$ 80$^{\circ} N$) as an example, the large scale forcing of water vapor is stiff (a) from step 242 to step 245 and is non-stiff (b) from step 56 to 59.  In stiff example, the magnitudes of the data in a small region quickly become dozens of times larger in a short period of time. However, the magnitudes remains essentially the same over time in Non-stiff example.  }
    \label{fig:stiff}
\end{figure*}

Fig.~\ref{fig:integration} shows the data interaction during simulation. Taking step $k$ as an example, a NeuroParametrization will accept input from the dynamics (Red Arrow in Fig.~\ref{fig:integration}), including large-scale states (i.e., water vapor~$q_{v,k}$ and temperature~$T_k$\,), surface pressure~($Ps_k$), and solar insolation~($Solin_k$) at the top of the atmosphere. It then predicts the corresponding output changing rate (tendency) of the moisture~$(\frac{\partial q_v}{\partial t})_k$ and dry static energy~$(\frac{\partial s}{\partial t})_k$ at each model level as $k+1$ step Dynamics' input (Black Arrow in Fig.~\ref{fig:integration}).

To enhance the regression capability of NeuroParametrization, we also introduce additional large-scale forcings variables (i.e., water vapor forcing~$(\frac{\partial q_v}{\partial t})_{l.s_k}$ and temperature forcing~$(\frac{\partial T}{\partial t})_{l.s_k}$\,) as additional input data, which are the sum tendencies from the previous dynamics and additional modules. Also, it is essential to include direct and diffuse downwelling solar radiation fluxes~$\Phi_{e,k}$ as output variables to force the coupled land surface model under the real land and ocean distribution. The inputs and outputs of NeuroParameterization are summarized in Table~\ref{tab:basic}. The shape and characteristics of these variables are described in detail in Sec.~\ref{sec:dataset}.
\begin{table}[htbp]
\renewcommand\arraystretch{1.5}
  \centering
  \caption{The dataset of input and output. The Input $X_k$ consists of 6 different physical information. The Output $Y_k$ consists of 3 different physical information.}
  \begin{adjustbox}{width=0.5\textwidth,center}
    \begin{tabular}{llll}
    \toprule
    Type  & \multicolumn{1}{c}{Full Name} & Channel Index & Dimension \\
    \midrule
    Input & state of water vapor at all levels ($q_{v,k}$)                    &   0:29  & 30,96,144 \\ 
    Input & state of temperature at all levels ($T_k$)                        &  30:59  & 30,96,144 \\
    Input & large scale forcing of temperature at all levels ($(\frac{\partial T}{\partial t})_{l.s_k}$ ) &  90:119 & 30,96,144 \\
    Input & large scale forcing of water vapor at all levels ($(\frac{\partial q_v}{\partial t})_{l.s_k}$)  &  60:89  & 30,96,144 \\
    Input & surface pressure ($Ps_k$)                                         & 120:121 & 1,96,144 \\
    Input & top-of-atmosphere solar insolation ($Solin_k$)                    & 121:122 & 1,96,144 \\
    \midrule
    Output & tendencies of water vapor at all levels ($(\frac{\partial q_v}{\partial t})_k$)  &  0:29  & 30,96,144 \\
    Output & tendencies of temperature at all levels ($(\frac{\partial s}{\partial t})_k$)    & 30:59  & 30,96,144 \\
    Output & 4 shortwave radiation fluxes reaching the surface ($\Phi_{e,k}$)                 & 60:64  &  4,96,144 \\
    \bottomrule
    \end{tabular}%
    \end{adjustbox}
  \label{tab:basic}%
\end{table}
\subsection{Fortran-Python Interface}
\label{sec:FPI}

Unlike other sub-grid parameterization schemes, the core component of NeuroParametrization is neural networks (NNs), which means that it is developed in C++ or Python. On the other hand, Fortran is the de facto standard of the scientific community, especially for climate modeling (i.e., CAM, SPCAM, and other GCMs). So, in developing NeuroGCM, one has to interface the machine learning components with complex source codes in climate modeling. It should be noted that we do not discuss Fortran-based machine learning frameworks here, as they are hardly ever widely used. 

Usually, in NeuroGCM simulation, NeuroParameterization should be called as a function. However, NeuroParametrization is written in the dynamically typed language Python. And it is well known that Fortran cannot call a python function directly. In our work, we developed the Fortran-Python Interface for data communication and synchronization control between NeuroParametrization and NeuroGCM. The main principle of the Fortran-Python Interface is to use system-level interfaces to communicate data between NeuroGCM and machine learning components (i.e., NeuroParametrization). 
Fig.~\ref{fig:architecture}(c) shows the structure of the Fortran-Python Interface in NeuroParametrization.
Based on the Fortran-Python Interface, Neuro-Parametrization can support all major machine learning frameworks, including PyTorch and Tensorflow, and effectively supports the development of the data-driven hybrid modeling \cite{hu2018squeeze,he2021blending,huang2021rethinking} community.

\section{Dataset}
\label{sec:dataset}

\subsection{Overview}

This section introduces in detail the input data and the output data. We summarize the physical information and the data shape of the dataset in Table~\ref{tab:basic}. As introduced in Sec.~\ref{sec:workflow}, the dynamics of NeuroGCM has a horizontal resolution of 1.9°×2.5°~(96*144) and 30 vertical levels. 

\begin{figure}[t]
    \centering
    \includegraphics[width=0.49\linewidth]{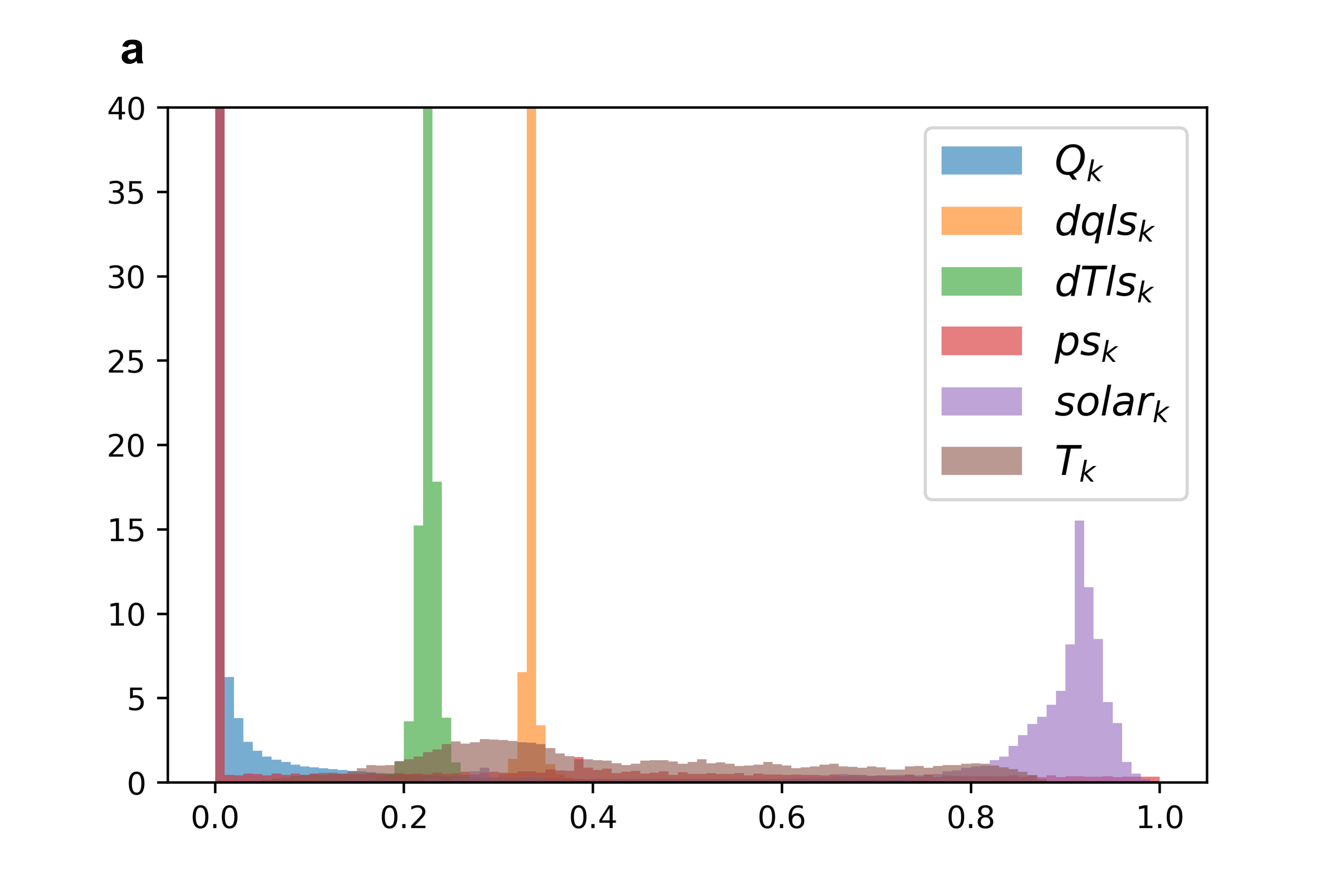}
    \includegraphics[width=0.49\linewidth]{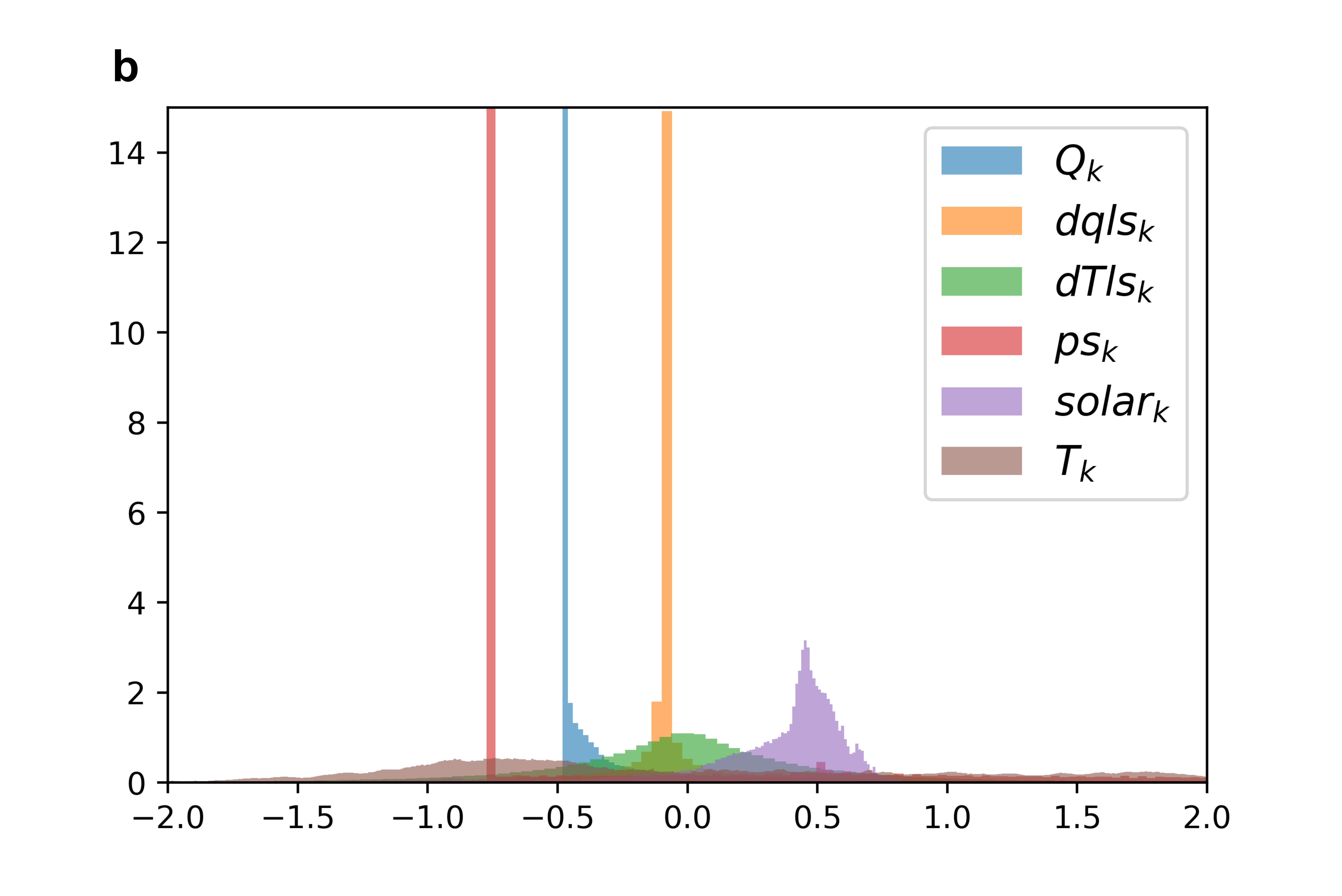}
    \caption{The normalization result of the input data. (a) max-min normalization; (b) normal normalization.}
    \label{fig:disti}
\end{figure}

\textbf{For the Input of the Dataset.} As shown in the Table~\ref{tab:basic}, the input $X_k$ consists of 6 different physical information, and the total dimension is (122, 96, 144), where 96 and 144 represent latitude and longitude, respectively. The large scale forcing of water vapor~($(\frac{\partial q_v}{\partial t})_{l.s_k}$) and large scale forcing of temperature~($(\frac{\partial T}{\partial t})_{l.s_k}$) are measured by vertical profile of water vapor and vertical profile of water vapor at two adjacent time steps. 

The data types of the last two channels in $X_k$ are surface pressure~($\textbf{Ps}_k$) and the top-of-atmosphere solar insolation~($\textbf{Solin}_k$). The shape of them are (1, 96, 144) .

\textbf{For the Output of the Dataset.} As shown in the Table~\ref{tab:basic}, the Output $Y_k$ consists of 3 different physical information, and the total dimension is (65, 96, 144). The first 60 channels represent the tendencies of water vapor~($(\frac{\partial q_v}{\partial t})_k$) and dry static energy~($(\frac{\partial s}{\partial t})_k$) due to the moist physics. The remaining channels of the $Y_k$ is the shortwave radiation fluxes reaching the surface.

To better demonstrate the physical meaning of the dataset, in Fig.~\ref{fig:datashow} in Appendix, we select the input and output data of a specific step and visualize its individual components.

\begin{figure*}[t]
    \centering
    \includegraphics[width=\linewidth]{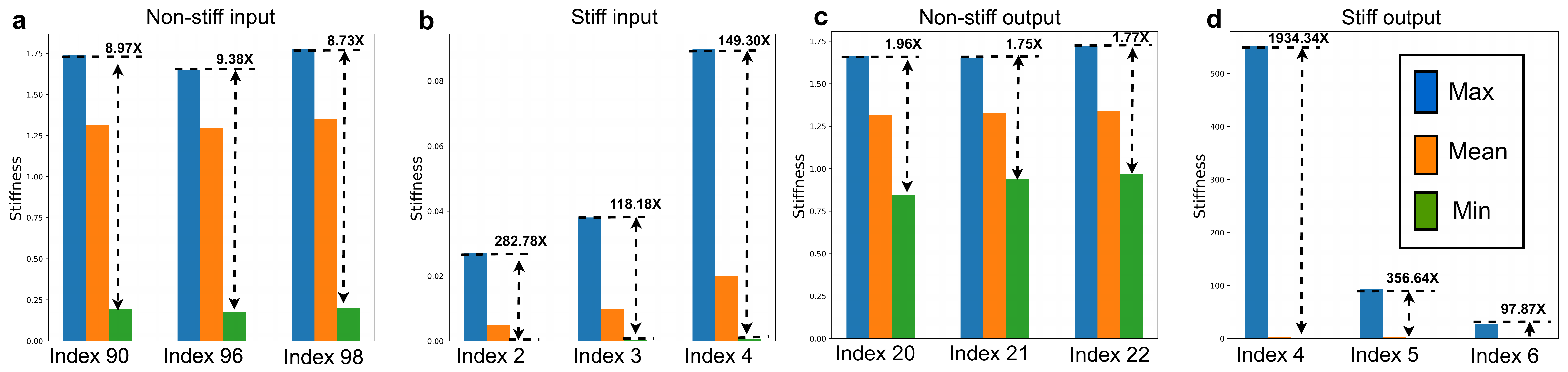}
    \caption{The ratio of the maximum and minimum of stiffness value. "Max", "Mean" and "Min" represent the maximum, mean and minimum values, respectively. "8.97x" denote the ratio of "Max" and "Min" is 8.97. }
    \label{fig:stiffnonstiff}
\end{figure*}

\subsection{Numerical characteristics}
\label{sec:numcha}
In this section, we analyze the numerical characteristics of the data and discuss the difficulties of data-driven prediction through these data.

\textbf{Heterogeneity.} From Fig.~\ref{fig:datashow} in Appendix, we can find that the magnitudes of the input data and output data in different channel index are very different, and such differences have a large impact on the data-driven prediction methods. To address such impact, some normalization methods of the data are generally performed. We normalize the data by max-min normalization and normal normalization, respectively, and the results are shown in Fig.~{\ref{fig:disti}}. We can notice that these two normalization methods can not eliminate the difference of magnitude in the data distribution. This difference can be seen as a feature bias and affects the training of the model, which would be a serious challenge for the data-driven approach.

\textbf{Stiffness.} Climate data tends to be elusive over time and is sensitive to initial values. Therefore climate problems are also often considered as chaotic. Chaotic system tends to have the problem of stiffness\cite{liang2022stiffnessaware}, i.e., drastic changes in a short period of time. As shown in Fig.~\ref{fig:stiff}(a), Taking the region of longitude (0$^{\circ} E\sim$ 140$^{\circ} E$) and latitude (0$^{\circ} N\sim$ 80$^{\circ} N$) as an example, the large scale forcing of water vapor undergoes a dramatic change in a very short period of time, where each step is 30 min. Specifically, when the step is 243 and 244, the magnitudes of the values in a small region become several times larger. 
However, when the step is 245, the magnitude of the large scale forcing of water vapor reverts back to that of step 242. When training on stiffness data, neural network methods can usually lead to an unstable solution\cite{kim2021stiff} or biased estimation, since the NN-based optimization has an implicit bias toward fitting a smooth function\cite{xu2019frequency,cao2019towards} with the fast decay in the frequency domain. Therefore, the researchers should pay more attention to get rid of the problem of such stiffness while designing the data-driven algorithms.

In fact, some of the data in the dataset are non-stiff, as shown in the Fig.~\ref{fig:stiff}(b). The large scale forcing of water vapor is relatively smooth with time, i.e., the magnitude of the data is maintained over time at e-11, which is friendly to the training of the model. Next, we explore the stiffness of different components of the data. For any given data $\{A_t^i\}$ with channel index $i$, we use stiffness value \cite{liang2022stiffnessaware}, i.e., Eq.~(\ref{eq:stiff}), to measure the stiffness.

\begin{equation}
    \frac{1}{\left\|{A_{t}}^{i}\right\|_{2}}\left\|\frac{{A_{t+1}}^{i}-{A_t}^{i}}{\Delta t}\right\|_{2}
    \label{eq:stiff}
\end{equation}

As seen from Eq.~(\ref{eq:stiff}), stiffness value is used to describe the short-term change rate of the data $\{A_t^i\}$. The greater the stiffness value, the greater the stiffness of the data at step $t$. In Fig.~\ref{fig:stiffnonstiff}, we select several cases for stiff and non stiff in the input data and output data. 
In Fig.~\ref{fig:stiffnonstiff}(a) and Fig.~\ref{fig:stiffnonstiff}(c), the ratio of the maximum and minimum of stiffness value is not large. However, in stiff case, the ratio is much larger than that in Fig.~\ref{fig:stiffnonstiff}(a) and Fig.~\ref{fig:stiffnonstiff}(c). For example, in Fig.~\ref{fig:stiffnonstiff}(d), when the channel index is 4, the ratio exceeds 1800, which means that there is severe stiffness in the data. However, this phenomenon is not negligible in studying climate data, so effective methods need to be designed to overcome this phenomenon for better data-driven prediction models.

\section{Experiment}

\subsection{Method} 
\label{sec:method}

This section introduces the implementation of NeuroParameterization (i.e., data-driven method) used in NeuroGCM in this paper, which supports long-term stable simulation. This implementation can be used as a baseline for other data-driven methods in NeuroGCM. 
Our implementation with deep neural networks mitigate the heterogeneous and stiff issues discussed in Sec.~\ref{sec:numcha}.

Specifically, as discussed in Section~\ref{sec:numcha}, the output data has well separated scales for each component and the phenomena of rapid change (stiffness). To better capture the pattern and enhance the training stability, we individually define an NN model to regress each component. As the output data comprises three components, three independent NN models are used with the same input size 122 and output size 30, 30, and 5 respectively. Besides, we employ the residual network (ResMLP) to regress the data. The structure is shown in Appendix, and ResMLP is a composition of linear transformation and nonlinear activation functions, and it is equipped with skip connections every two layers. 

To increase the generalization of the trained models, we augment the data by adding the Gaussian noise $\mathcal{N}(0,0.001)$ to both the input data and output data. 


We propose two metrics for the approximation ability of the data-driven methods (i.e, NeuroParameterizations) and the simulation stability of NeuroGCM.

\subsection{The Metrics for Machine Learning}

In Section~\ref{sec:platform}, we mentioned that the whole framework requires NeuroParameterization modules and dynamics modules to interact with each other. However, in the neural network training process, if the entire interactive process is considered, the training process will be very slow, and the backpropagation of the gradient can not be carried out effectively.

For this reason, we have to consider alternative evaluation metrics for the training of NeuroParameterization. Specifically, we use mean square error~(MSE) to be the loss of our baseline method, i.e., $MSE = \frac{1}{N}||Y - Y_{\textbf{label}}||^2$, where $N$ is the number of the training data and $Y_{\textbf{label}}$ denotes the ground truth output data. In addition, we can also take $RMSE$ as the loss as shown in Eq.~(\ref{qe:rmse})
\begin{equation}
    RMSE = \sqrt{\frac{1}{N}||Y - Y_{\textbf{label}}||^2}.
    \label{qe:rmse}
\end{equation}

Since the variation of various physical quantities in the vicinity of the equator is very drastic. This makes the performance of the model heavily dependent on the training of the model in the region near the equator. Therefore, using Eqn.~(\ref{qe:rmse}), we consider a new metric $RMSE_{\text{Equator}}$ for machine learning, \textit{i.e.}, the $RMSE$ in range of [15$^{\circ}$ N, 15$^{\circ}$ S]. 

\begin{table}[htbp]
  \centering
  \caption{The results of the metrics for machine learning. ResMLP is the model shown in Appendix and MLP is ResMLP without skip connections.}
  \resizebox{0.8\columnwidth}{!}{
    \begin{tabular}{lcrr}
    \toprule
    \textbf{Model} & \textbf{Channel Index} & \multicolumn{1}{l}{\textbf{  $RMSE$} } & \multicolumn{1}{l}{\textbf{$RMSE_{\text{Equator}}$}} \\
    \midrule
    MLP   & 0:29  & 4.1$\times 10^{-5}$       & 1.3$\times 10^{-4}$  \\
    MLP   & 30:59 & 3.6$\times 10^{-5}$       & 9.8$\times 10^{-5}$  \\
    MLP   & 60:64 & 2.9$\times 10^{-4}$       & 3.3$\times 10^{-4}$  \\
    \midrule
    ResMLP & 0:29  &  3.9$\times 10^{-5}$     & 1.2$\times 10^{-4}$   \\
    ResMLP & 30:59 &  2.9$\times 10^{-5}$     & 7.6$\times 10^{-5}$  \\
    ResMLP & 60:64 &  1.3$\times 10^{-4}$     & 1.4$\times 10^{-4}$  \\
    \bottomrule
    \end{tabular}%
    }
  \label{tab:r2}%
\end{table}%

We use the dataset proposed in Section~\ref{sec:dataset} to verify the performance of our baseline model in Section~\ref{sec:method}. The results are shown in Table~\ref{tab:r2}. From a machine learning perspective, even though $RMSE$ and $RMSE_{\text{Equator}}$ look small, that doesn't mean the model will perform well when it's actually used. Therefore we also need to consider the effect when simulating the actual climate change.

\subsection{The Metrics for Simulation}

\begin{figure*}[t]
    \centering
    \includegraphics[width=0.8\linewidth]{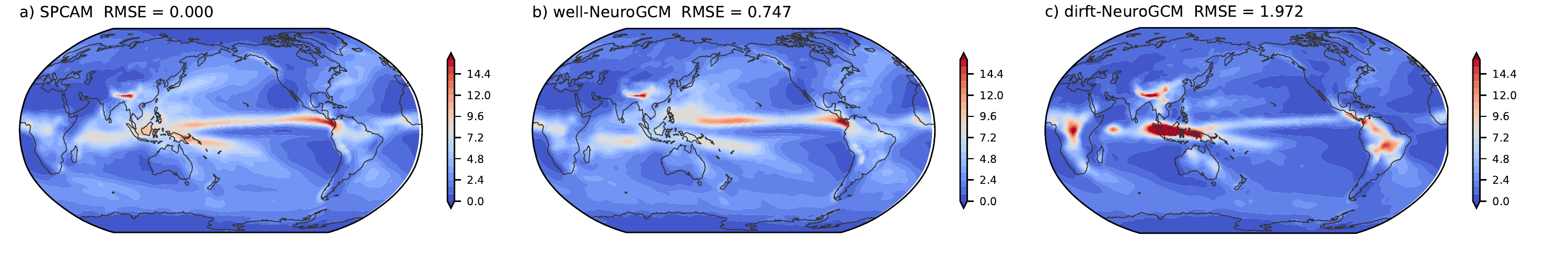}
    \caption{The 5-year averaged global rainfall intensity. (a) is the global rainfall intensity of SPCAM as a baseline (black curve in Fig. 10); (b) corresponds to the well-simulated NeruoGCM in Fig. 10 (green curve); (c) corresponds to the drift NeuroGCM in Fig. 10 (orange curve).}
    \label{fig:precp}
\end{figure*}

NeuroGCM is a hybrid model that integrates numerical simulation with deep learning. As shown in Fig.~\ref{fig:integration}, at each step of NeuroGCM integration, the input of the deep learning model is influenced by the output of the previous step. The errors generated by each inference of the deep learning model accumulate in the continuous integration of NeuroGCM, making NeuroGCM's essential states anomalous and affecting the results of climate prediction. 

\begin{figure}[t]
    \centering
    \includegraphics[width=0.705\linewidth]{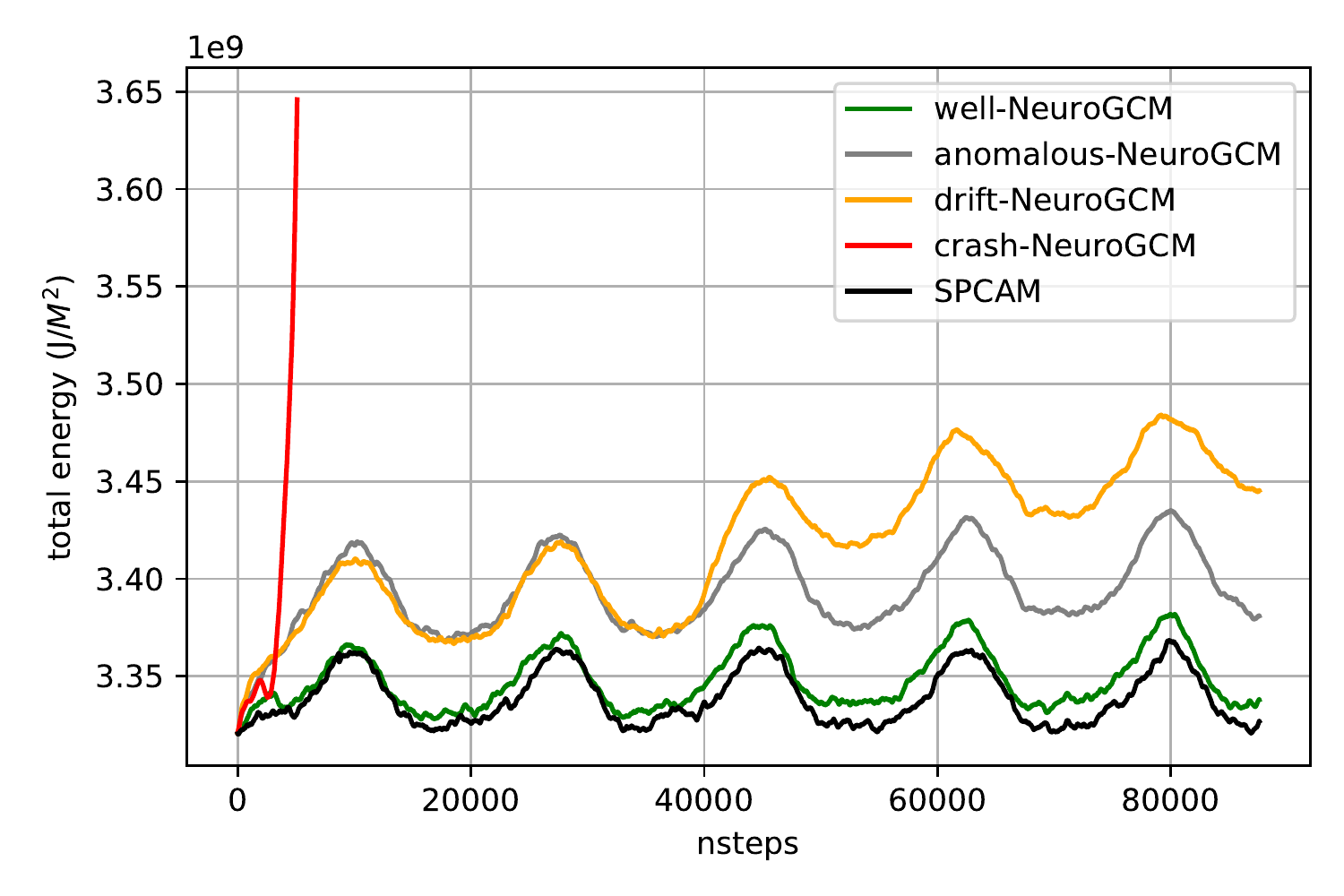}
    \caption{The global average total energy for all cases of NeuroGCMs and SPCAM.}
    \label{fig:totalenergy}
\end{figure}

Global rainfall intensity is an essential metric for assessing GCMs~\cite{randall2007climate}. Fig.~\ref{fig:precp} shows the 5-year averaged global rainfall intensity in the NeuroGCM long-term simulations with different NeuroParameterizations. (Note, These NeuroParameterizations have similar RMSEs.) The well-simulated NeuroGCM (Fig.~\ref{fig:precp} (b)) can resemble the target SPCAM simulated rainfall, while another NeuroGCM performs oddly with large anomalous precipitation found in the simulation results (Fig.~\ref{fig:precp} (c)), marked by 2.64 times root mean square error.

Furthermore, abnormal simulation results may even cause the hybrid modeling to collapse. The global average total energy is an essential indicator in GCMs, with a unit of $J/m^2$~\cite{enting2018metrics}. We can use it to visualize the stability of NeuroGCMs during the simulation. Fig.~\ref{fig:totalenergy} shows the global averaged total energy via time sequence in the NeuroGCMs (with similar RMSEs (i.e., test loss) for their NNs) and the SPCAM. 
The average total energy in the target SPCAM shows seasonal fluctuations without a systematic trend to increase or decrease, and the global averaged total energy of the well-simulated NeuroGCM (the green line) is very close to that of the SPCAM as a reference. In contrast, some other NeuroGCMs show deviated or drifting total energy, and some even crash with soaring energy curves. For instance, the crashed NeuroGCM (red curve in Fig.~\ref{fig:totalenergy}) only ran 106 days (about 5000 steps), and we could not even get the multi-year averaged global rainfall intensity as in Fig.~\ref{fig:precp}. 
Therefore, a quantifiable metric that measures the impact of different NeuroParameterizations on the stability of NeuroGCM simulations is as important as the test loss (i.e., RMSE for neural networks) when evaluating the totality of NeuroGCM.

For this purpose, we introduce the pseudo-Radiative forcing to measure the systematic changing rate of the total energy in the NeuroGCM simulations. It is well known that Radiative forcing is an important indicator of global climate change~\cite{stocker2014climate} and is used to quantify the impact of greenhouse gases such as carbon dioxide and methane on the total energy of the atmosphere, which is measured in $W/m^2$. Similarly, simulation errors in GCMs due to biases can cause the pseudo-Radiative forcing, which manifests itself as climate drift or collapse.

The pseudo-Radiative forcing $\kappa$ is the slope of the energy curves, calculated by the least square method as 
\begin{equation}
    \kappa = \frac{\sum_{i=1}^{T}\left(t_{i}-\bar{t}\right)\left(e_{i}-\bar{e}\right)}{\sum_{i=1}^{T}\left(t_{i}-\bar{t}\right)^{2}},
    \label{eq:Prf}
\end{equation}
where $t_i$ and $e_i$ denote the $i^{\rm th}$ step and total energy. $\bar{t}$ and $\bar{e}$ are $\frac{1}{T}\sum_{i=1}^Tt_i$ and $\frac{1}{T}\sum_{i=1}^Te_i$. The closer the $\kappa$ is to 0, the more stable the simulation is. 


Table~\ref{tab:rf} shows the pseudo-Radiative forcings $\kappa$ of CAM, SPCAM, and NeuroGCMs, including crash-NeuroGCM, well-simulated NeuroGCM, and anomalous simulated NeuroGCM. For the well-simulated NeuroGCM, $\kappa$ is 0.015579 $W/m^2$, which is in the order of SPCAM and CAM.

\begin{table}[htbp]
  \centering
  \caption{The results of Radiative forcing for all models in the Metrics for Simulation. Let $T = 87600$. There are $365\times 5\times 24 \times 2 = 87600$ steps (30 minutes) in 5 years.}
  \resizebox{0.8\columnwidth}{!}{
    \begin{tabular}{lc} 
    \toprule
    GCMs                              & pseudo-Radiative forcing $\kappa$   \\
    \toprule
    \textbf{SPCAM} (baseline)         & \textbf{0.008147}                   \\
    CAM5                              & 0.010424                            \\
    \textbf{well-NeuroGCM}            & \textbf{0.015579}                   \\
    anomalous-NeuroGCM                & 0.166282                            \\
    drift-NeuroGCM                    & 0.688410                            \\
    crashed-NeuroGCM                  & 23.81831                            \\
    \bottomrule
    \end{tabular}
    }
  \label{tab:rf}%
\end{table}

\subsection{Speed}

We test the computing performance of NeuroParametrization and NeuroGCM separately, using NeuroGCM's learning target SPCAM as a control group to evaluate their potential to improve computational performance. We did not comprehensively make parallel performance optimization on NeuroParametrization and NeuroGCM. Therefore, it should be easy to obtain higher speeds \cite{huang2022accelerating}.

For the evaluation, we used 192 Intel CPU Cores (from Intel(R) Xeon(R) Gold 6132 CPUs) to run SPCAM and NeuroGCM, with NeuroGCM using an additional Nvidia Tesla V100 GPU to accelerate the neural network inference.

SuperParametrization (CRM and its radiation process) is the most time-consuming process in SPCAM's simulation. In NeuroGCM, the corresponding process is NeuroParametrization, which includes GPU-accelerated neural network inference and data communication. In NeuroClim, NeuroParametrization was \textbf{41x} faster than SuperParametrization, and NeuroGCM was \textbf{35x} faster than SPCAM (see Table~\ref{tab:speed}). 

\begin{table}[htbp]
  \centering
  \caption{The speed benchmark for NeuroClim includes parameterization schemes (the table above) and time for each simulation step (the table below).}
  \resizebox{0.6\columnwidth}{!}{
    \begin{tabular}{lc} 
    \toprule
    Parameterization Schemes    & Time (s) \\ 
    \toprule
    SuperParameterization       & 16.2792  \\ 
    NeuroParameterization       & 0.3923   \\ 
    \bottomrule
                                           \\
    \toprule
    GCMs                        & Time (s) \\ 
    \toprule
    SPCAM                       & 17.0997  \\ 
    NeuroGCM                    & 0.4839   \\
    \bottomrule
    \end{tabular}
    }
  \label{tab:speed}%
\end{table}

\section{Related Works} 
\label{sec:related}
In the last eight years, data-driven hybrid modeling have been developed with more advanced ML architectures under more complex application scenarios. The researchers initially used machine learning algorithms to make one-time predictions with the input from the host GCMs with the simple oceanic surface. \citet{krasnopolsky2013using} trained a neural network ensemble with the simulation results from a regional CRM during the TOGA-COARE period in the tropical western Pacific and achieved accurate one-time predictions on rainfall. \citet{gentine2018could} used training data from SPCAM simulations under the ``Aqua-planet'' setting (an ideal global model with all surfaces covered with prescribed sea surface) to develop deep neural networks that can emulate convection and clouds and radiation processes with the desired offline accuracy. Then researchers put the ML predictions back to the coarse resolution primary numerical method and continued integrating the hybrid modeling, enabling a true hybrid simulation but still under an ideal Aqua-plane setting. \citet{rasp2018deep} improved the neural network of \citet{gentine2018could}, allowing for the first time a hybrid run of a deep neural network subgrid parameterization scheme with GCMs in the Aqua-plane setting (i.e., hybrid modeling). However, this deep learning parameterization still contained uncertainties~\cite{rasp2020coupled} that the neural network architecture and training data seriously affected the duration of stable hybrid runs. Nonetheless, \citet{yuval2020stable} and \citet{yuval2021use} developed long-term stable ML-based parameterizations with separated predictions of different processes in convection and clouds.

Apart from the aqua-planet settings, some researchers recently attempted to implement hybrid modeling under the real-world setting. \citet{han2020moist} emulated SPCAM simulations with a 1-D ResNet in the real-world setting and single-column versioned GCM with desirable results. \citet{mooers2021assessing} used AutoML to select the best deep learning model and hyperparameters. They found that the ability of machine learning to fit cloud and convective processes are weakened since the complexity of convection increases significantly when the realistic land and sea distributions are involved.
Achieving multi-year stable hybrid simulations under the real-world setting is challenging toward radicalizing hybrid modeling. The first success is \citet{wang2021stable}. They used a group of residual multilayer perceptrons to achieve long-term stable simulations of hybrid modeling under realistic land-sea distributions. They obtained reasonable climate simulation results but with unneglectable biases.

\section{Conclusion}


We propose a hybrid modeling framework NeuroClim which comprises an open-source platform (NeuroGCM), a dataset for building data-driven methods, and two metrics to evaluate hybrid modeling quantitatively. Moreover, this paper provides a baseline data-driven parameterization scheme, which supports long-term stable climate simulation.
In the future, our work can prompt both the machine learning community and the climate modeling community to address impactful challenges in the AI-Climate intersection. Such as the interpretability and stability of hybrid modeling and the evaluation of climate prediction skills for different regions.

\clearpage
\appendix
\section{The details of the equations for the numerical methods}
\label{numerical_eqn}

Here we introduce the equations involved in this work, including the numerical solution of large-scale processes in Dynamics and convection and clouds in SPCAM.

In climate models, large-scale motions are also called grid-scale motions. Specifically, grid-scale motions $\bar{u}_{i}$ for three dimensions ($i=1, 2, 3$) and scalar variables $\bar{A}$ such as dry static energy and water vapor are shown as:
\begin{small}
\begin{align}
    &\frac{\partial \bar{u}_{i_{1,2}}}{\partial t}+\frac{1}{\bar{\rho}} \frac{\partial}{\partial x_{j}}\left(\bar{\rho} \bar{u}_{i_{1,2}} \bar{u}_{j}\right)+\frac{1}{\bar{\rho}} \frac{\partial \bar{p}}{\partial x_{i_{1,2}}}=\left(\frac{\partial \bar{u}_{i_{1,2}}}{\partial t}\right)_{sub} \tag{1} \label{eqn:lsm_1}\\
    &-\frac{1}{\bar{\rho}} \frac{\partial \bar{p}}{\partial x_{i_{3}}}=g \tag{2} \label{eqn:lsm_2}\\
    &\frac{\partial}{\partial x_{j}}\left(\bar{\rho} \bar{u}_{j}\right)=0 \tag{3} \label{eqn:lsm_3}\\
    &\frac{\partial \bar{A}}{\partial t}+\frac{1}{\bar{\rho}} \frac{\partial}{\partial x_{j}}\left(\bar{\rho} \bar{A} \bar{u}_{j}\right)=\left(\frac{\partial \bar{A}}{\partial t}\right)_{sub} \tag{4} \label{eqn:lsm_4}
\end{align}
\end{small}
where $\bar{\rho}$ is the grid air density, $g$ is the gravity constant, (\ref{eqn:lsm_1}) are the momentum equations horizontally, (\ref{eqn:lsm_2}) is vertical equation, assumed hydrostatic equilibrium, (\ref{eqn:lsm_3}) is the continuity equation and (\ref{eqn:lsm_4}) is the equation for scalar variables. Therefore, the grid-scale dynamics, characterized by slow vertical motion and zero vertical acceleration due to the hydrostatic equilibrium, need the tendencies from subgrid processes like $\left(\frac{\partial \bar{u}_{i_{1,2}}}{\partial t}\right)_{sub}$ and $\left(\frac{\partial \bar{A}}{\partial t}\right)_{sub}$.

For SPCAM, convection and clouds inside each coarse grid are resolved by the thermal dynamics (equation \ref{eqn:crm_1}, \ref{eqn:crm_2}, \ref{eqn:crm_3}). This unique subgrid parameterization embeds a sub-numerical model (i.e., CRM) with high resolution inside each coarse grid. In this research, the 2-D CRM contained 32 grid points in the zonal direction and 30 vertical levels shared with the grid-scale dynamics. Likewise, the equations for the thermal dynamics of the 2-D CRM are listed as:

\begin{small}
\begin{align}
&\frac{\partial u_{i}}{\partial t}=-\frac{1}{\rho} \frac{\partial}{\partial x_{j}}\left(\rho u_{i} u_{j}+\tau_{i j}\right)-\frac{1}{\bar{\rho}} \frac{\partial p^{\prime}}{\partial x_{i}}+\delta_{i_{3}} B+\left(\frac{\partial \bar{u}_{i}}{\partial t}\right)_{l.s.} \tag{5} \label{eqn:crm_1}\\
&\frac{\partial}{\partial x_{j}}\left(\rho u_{j}\right)=0 \tag{6} \label{eqn:crm_2}\\
&\frac{\partial A}{\partial t}=-\frac{1}{\rho} \frac{\partial}{\partial x_{j}}\left(\rho A u_{j}+F_{A, i}\right)+S_{A}+\left(\frac{\partial \bar{A}}{\partial t}\right)_{l.s.} \tag{7} \label{eqn:crm_3}
\end{align}
\end{small}
where $\bar{u}_{i}$, $A$, and $\rho$ are similar to those in equation (1) to (4) but for the subgrid 2-D CRM model.

Equation (\ref{eqn:crm_1}) is corresponding to (\ref{eqn:lsm_1}) and (\ref{eqn:lsm_2}), in which the vertical acceleration is majorly affected by the buoyancy $B=-g \frac{T^{\prime}}{T}\left(T=\bar{T}+T^{\prime}\right)$ and the perturbed pressure gradient $\frac{1}{\bar{\rho}} \frac{\partial p^{\prime}}{\partial x_{i 3}}\left(p=\bar{p}+p^{\prime}\right)$. The resolved thermal dynamics in the CRM are far more nonlinear and non-balanced than the grid-scale governing equation, involving inertia, frictions $\tau_{i j}$, and diffusions $F_{A, i}$. Also, the CRM needs the grid-scale forcings $\left(\frac{\partial \bar{u}_{i}}{\partial t}\right)_{\text {l.s.}}$ and $\left(\frac{\partial A}{\partial t}\right)_{l.s.}$\,as input to keep up with the states of the grid points, and outputs the temporal and spatial averaged changing rate as subgrid tendencies $\frac{\partial \overline{A}}{\partial t}=\left(\frac{\partial \bar{A}}{\partial t}\right)_{sub}$.

\section{The details of Training}

\begin{table}[htbp]
  \centering
  \caption{The details of the neural network shown in Fig.~\ref{fig:resmlp}. The ResMLP is a composition of several blocks. There are two linear transformation and nonlinear activation functions in each block. ``lr'' denotes the original learning rate for the Adam optimizer. ``node size'' means the number of neuron in each linear transformation.}
    \begin{tabular}{ll}
    \toprule
    Type                & Details \\
    \midrule
    Activation function & ReLU    \\
    Epoch               & 50      \\
    Number of blocks    & 7       \\
    lr                  & 0.001   \\
    lr strategy         & coslr   \\
    node size           & 512     \\
    Optimizer           & Adam    \\
    Batch size          & 1024    \\
    \bottomrule
    \end{tabular}%
  \label{tab:hyperparameter}%
\end{table}%

\begin{table}[htbp]
  \centering
  \caption{The maximum and minimum in input data and output data with different channel index. We use max-min normalization to normalize the data, i.e., $\text{Normed data} = (\text{data} - \text{min})/(\text{max} - \text{min}) \times 2 -1 $.}
    \begin{tabular}{llll}
    \toprule
    Type   & Channel Index & Max     & Min      \\
    \midrule
    Input  & 0:29          & 0.0238  & 0        \\
    Input  & 30:59         & 323     & 159      \\
    Input  & 60:89         & 2.13e-6 & -2.13e-6 \\
    Input  & 90:119        & 3.89e-3 & -3.89e-3 \\
    Input  & 120:121       & 1412    & 0        \\
    Input  & 121:122       & 105782  & 59928    \\
    Output & 0:29          & 3.11e-6 & -3.11e-6 \\
    Output & 30:59         & 3.63    & -3.63    \\
    Output & 60:64         & 1412    & 0        \\
    \bottomrule
    \end{tabular}%
  \label{tab:datanorm}%
\end{table}%

\begin{figure}[H]
    \centering
    \includegraphics[width=0.8\linewidth]{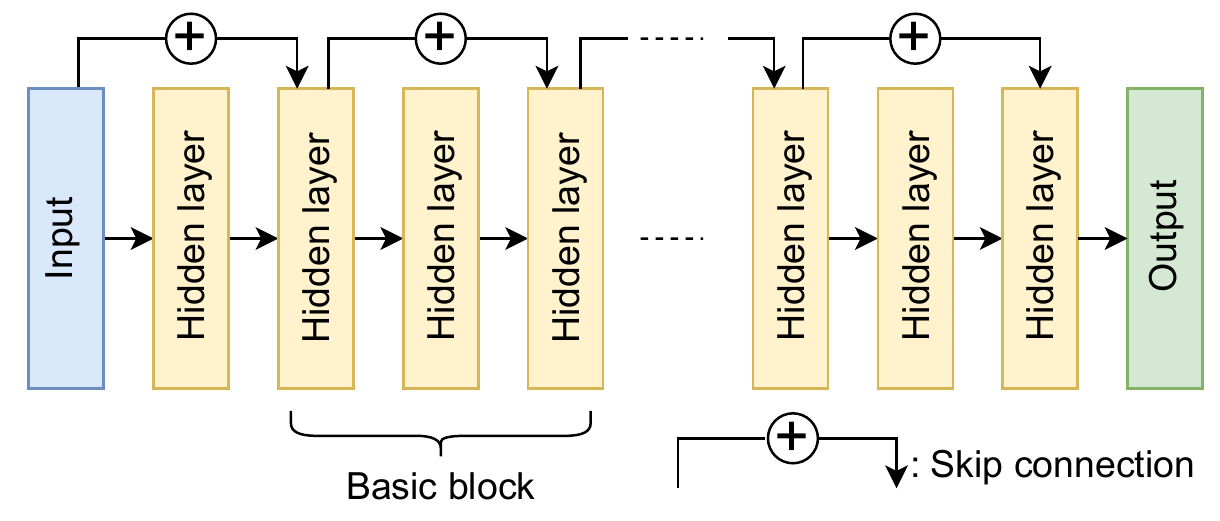}
    \caption{The structure of the ResMLP.}
    \label{fig:resmlp}
\end{figure}
\begin{figure*}
    \centering
    \includegraphics[width=1.0\linewidth]{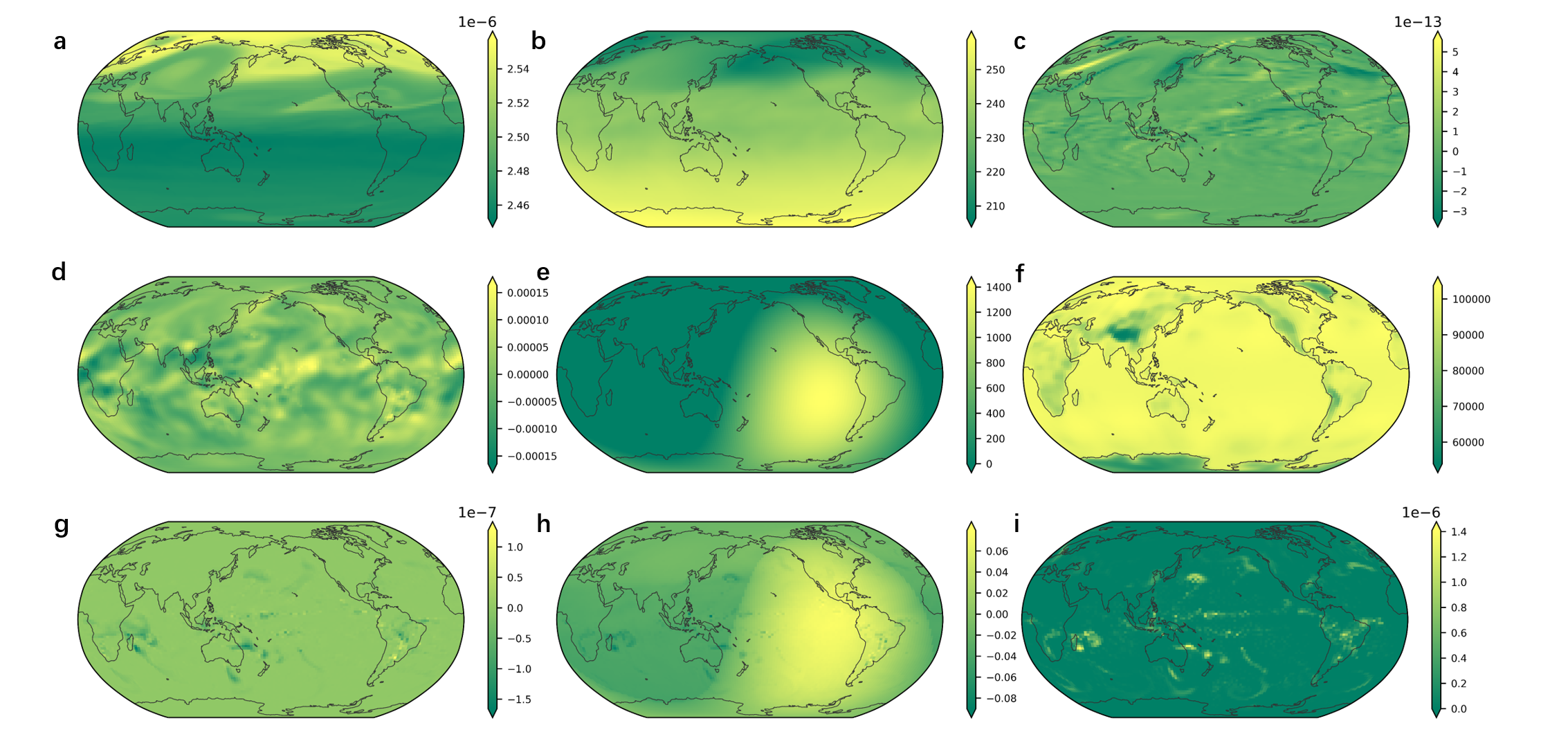}
    \caption{Visualization of the data. The subplots in the first two rows represent each of the six components of the input. (a) The vertical profile of water vapor ($\textbf{q}_{v,k}$); (b) The profile of temperature ($\textbf{T}_k$);  (c) The large scale forcing of water vapor ($(\frac{\partial q_v}{\partial t})_{l.s_k}$); (d) The large scale forcing of temperature~($(\frac{\partial q_v}{\partial t})_{l.s_k}$); (e) Surface pressure~($\textbf{Ps}_k$); (f) The top-of-atmosphere solar insolation~($\textbf{Solin}_k$). The subplots in the last row represent each of the three components of the output. The tendencies of (g) water vapor~($(\frac{\partial q_v}{\partial t})_k$) and (h) dry static energy~($(\frac{\partial s}{\partial t})_k$); (i) The shortwave radiation fluxes reaching the surface.}
    \label{fig:datashow}
\end{figure*}




\end{document}